\documentclass[twocolumn,showpacs,preprintnumbers,amsmath,amssymb]{revtex4}
\usepackage{graphicx}
\usepackage{dcolumn}
\usepackage{bm}
\def\bea {\begin{eqnarray}}
\def\eea {\end{eqnarray}}

\def\be {\begin{equation}}
\def\ee {\end{equation}}

\begin{document}
\title{Multiplicity distributions in the forward rapidity region in proton-proton collisions at the Large Hadron Collider}
\author{Premomoy Ghosh}
\email{prem@vecc.gov.in}
\author{Sanjib Muhuri}
\email{sanjibmuhuri@vecc.gov.in}
\medskip
\affiliation{Variable Energy Cyclotron 
Centre, 1/AF Bidhan Nagar, Kolkata 700 064, India}            
\date{\today}

\begin{abstract}
Measured multiplicity distributions of primary charged particles produced in the forward rapidity region of the $proton-proton$ ($pp$)
collisions at the centre-of-mass energy, $\sqrt {s}$ = 7 TeV at the Large Hadron Collider (LHC) have been analyzed in terms of the Negative 
Binomial Distribution (NBD) function. Like the multiplicity distributions in the mid-rapidity region for the $pp$ collisions at $\sqrt {s}$ = 
7 TeV, the distributions for the minimum bias events in the forward region also are better described with the superposition of two-NBDs, as proposed 
by a two-component model of particle production from two processes, the "$soft$" and the "$hard$". However, the multiplicity distribution 
for the "hard-QCD" events in a large pseudorapidity window does not oblige the two-component model.   
\end{abstract}

\maketitle
\section{Introduction}
The major experiments at the Large Hadron Collider \cite {ref01}, depending on specific physics requirements, have detector setups of 
different geometrical acceptance for detecting several kinds of particles in different kinematic ranges. Beside the specific physics goals,
all these detector setups facilitate study of the physics of the collisions, in general, by implicitly recording information on particle productions in 
terms of a few basic observables. Sometimes, comparisons of data recorded in different acceptance of detectors of these experiments could 
provide insights to the particle production mechanisms in different phase-space of collisions, due to different kinematic origin. In this respect, 
out of the four major experiments at the LHC, the Large Hadron Collider Beauty (LHCb) experiment has a unique standing. While A Large 
Ion Collider Experiment (ALICE), the Compact Muon Solenoid (CMS) experiment and A Toroidal LHC Apparatus (ATLAS) experiment 
primarily address the mid-rapidity physics by measuring majority of the produced charged particles in the mid-rapidity region, the LHCb 
setup allows measurement of charged particles in the forward rapidity region, facilitating the study of forward physics.
 
The multiplicity distribution of primary charged particles produced in collisions is one of the most basic observables, characterizing the 
final states of multi-particle production process in a high energy physics experiment. All the LHC experiments have measured \cite {ref02, 
ref03, ref04, ref05, ref06} multiplicity distributions in proton-proton ($pp$) - collisions at the available LHC energies in different kinematic 
ranges and for different classes of events. In the context of the present work, these LHC experiments, in spite of the differences in detector 
acceptance, have a common observation - the multiplicity distributions of produced particles at the new LHC energies have been found 
to be underestimated by several of the standard event generators / models (like PYTHIA, PHOJET etc.) in use. This observation has made 
the study of multiplicity distribution at LHC energies all the more interesting. 

\section{Objective}
In this article, we analyze the primary charged particle multiplicity distributions in the forward rapidity region in proton-proton ($pp$) - collisions at 
$\sqrt {s}$ = 7 TeV, as measured by the LHCb experiment at LHC, in terms of parameters of the Negative Binomial Distribution(NBD) function. 
The two-parameter NBD function, as given below in Eq. - (1) played a significant role in describing multiplicity distributions of produced charged
particles in the mid-rapidity region in $pp$ (and $p \bar p$) collisions for a wide range of the centre-of-mass energy, including $\sqrt {s}$ = 7 TeV. 

\begin{equation}
  P(n,\langle n \rangle, k) = \frac{\Gamma(k+n)}{\Gamma (k)\Gamma(n+1)}\left[\frac{\langle n \rangle}{k+\langle n \rangle}\right]^n \times \left[\frac{k}{k+\langle n \rangle}\right]^k
\end{equation}  
where $\langle n \rangle$ is the average multiplicity and the parameter $k$ is related to dispersion $D$, ($D^2 = \langle n^2 \rangle - \langle n \rangle^2$) 
by
\begin{equation}
\frac {D^2}{\langle n \rangle^2} = \frac{1}{\langle n \rangle} + \frac {1}{k}
\end{equation}  

The NBD function could describe the charged particle multiplicity distributions in $proton-antiproton$ ($p\bar p$) collisions at $\sqrt {s}$ = 540 
GeV at the Super Proton Synchrotron (SPS) \cite {ref07} at CERN in the full pseudorapidity ($\eta$) space as well as in limited pseudorapidity 
intervals (for high momentum low mass particles, the rapidity can be approximated to the pseudorapidity,  $\eta = -ln[tan(\theta/2)]$, where  
$\theta$ is the polar angle of the particle with respect to the counterclockwise beam direction). At $\sqrt {s}$= 900 GeV SPS energy, 
however, the single NBD function could describe the data only for small pseudorapidity intervals at the mid-rapidity region. With the appearance 
of sub-structures in multiplicity distributions at higher energies and in larger pseudorapidity intervals, the weighted superposition or 
convolution of more than one function including one NBD function,  as proposed by several models, \cite{ref08,ref09,ref10,ref11} representing 
more than one source or process of particle productions could explain the data better. Such sub-structure in SPS data at $\sqrt {s} =$ 900 
GeV and in Tevatron data at $\sqrt {s} =$ 1.8 TeV  \cite {ref12} could be well explained by weighted superposition of two NBD functions \cite{ref09}, 
as given by Eq. - 3. The multiplicity distributions of primary charged hadrons in Non-Single Diffractive (NSD) events in $pp$ - collisions at $\sqrt {s}$ = 
7 TeV in the mid-rapidity region also could be well explained \cite {ref13} by the two-NBD function.
\begin{eqnarray}
&&P_{n}(\sqrt {s},\eta_{c}) = \alpha_{soft}(\sqrt {s}) \nonumber\\
&&P_{n}[\langle n \rangle_{soft}(\sqrt {s},\eta_{c}),{k}_{soft}(\sqrt {s},\eta_{c})] + [1 - \alpha_{soft}(\sqrt {s})] \nonumber\\ 
&&P_{n}[\langle n \rangle_{semihard}(\sqrt {s},\eta_{c}),{k}_{semihard}(\sqrt {s},\eta_{c})]
\end{eqnarray}
where $\alpha_{soft}$ is the fraction of "soft'' events and is a function of $\sqrt {s}$ only. The other parameters, functions of both, the $\sqrt {s}$ and 
the $\eta_{c}$, have usual meanings as described for Eq. - (1) with suffixes in parameters indicating respective components.

At this point, discussing other models or approaches of multi-particle production involving NBDs would be relevant. The framework of the weighted 
superposition mechanism of different classes of events has been extended from the two-component to a three component model \cite {ref14} for 
explaining possible new physics at LHC at $\sqrt {s}$ = 14 TeV. The third component would attribute to the eventual new class of high multiplicity 
events which would be manifested by the appearance of a new elbow structure' in the tail of multiplicity distribution of the $pp$ collisions at the 
highest planned centre-of-mass energy at the LHC. So far, the multiplicity distributions for the $pp$ collisions up to $\sqrt {s}$ = 7 TeV are available
and these distributions have no such new structure, in the tail of the distributions, which calls for application of the model. A very recent theoretical  
approach \cite {ref15}, following Glasma flux tube model, has shown that the multiplicity distribution of multi-particle 
production could be described by convolution of a number of NBD functions as a natural consequence of several impact parameters of 
the collisions. The model reproduces the multiplicity distributions data of $pp$ collisions in small pseudorapidity window ($|\eta|<$ 0.5) 
at the LHC energies. The scope of the present work is, however, restricted to the analysis of the LHCb data in terms of a single NBD and a 
superposition of two NBDs, as prescribed by the two-component model of Ref. \cite {ref09}. 

According to the two-component model of Ref.- \cite{ref09}, the multiplicity distribution of hadronic collisions can be explained by weighted 
superposition of two NBDs, representing two classes of events, ``semihard - events with minijets or jets'' and ``soft - events without minijets or jets''. 
It is note worthy that the "semihard" events involving hard parton-parton scatterings (due to high momentum transfer) resulting in QCD jets of
high transverse momentum above a certain threshold is also referred to as "hard-QCD" events.

The LHCb experiment has measured \cite {ref06} multiplicity distributions of primary charged particles produced in the $pp$ collisions at 
$\sqrt {s}$ =  7 TeV in the pseudo rapidity coverages, $-2.5 < \eta < 2.0$ and $2.0 < \eta < 4.5$ for two classes of events: the minimum bias 
and the hard-QCD. The hard-QCD events were chosen out of the minimum bias events by selecting events with at least one particle with 
transverse momentum greater than 1 GeV/c. The multiplicity distributions for both the event-classes were measured for small pseudo rapidity 
windows of width $\eta_{c} = 0.5$ scanning over the $\eta$ - range of the detector coverage as well as for the  wide $\eta$ - window, $\eta_{c} = 
2.5$ ($2.0 < \eta < 4.5$). We analyze these distributions in the forward-rapidity in terms of the NBD that has been successful in describing the 
mid-rapidity data. We discuss the results, comparing with observations in similar analysis of data at the same $\sqrt {s}$ at the mid-rapidity region. 
For the mid-rapidity region, we consider the distributions, measured by the CMS experiment \cite {ref04}, as there exists \cite {ref13} similar 
phenomenological study of the CMS data in terms of the NBD-formalism.

\section{Analysis and Discussions}
\label{}
We fit the multiplicity distributions of primary charged particles, as measured \cite{ref09} by the LHCb, in the five pseudorapidity 
windows of width $\eta_{c} = 0.5$ in the $\eta$ - range, $2.0 < \eta < 4.5$ for the minimum bias events  with the NBD function as given in 
Eq. -1. The Table - ~\ref{tab:NBD_minbias} contains the values of the parameters obtained by the best fits, along with the the corresponding 
values of $\chi^2/ndf$. 
\begin{table}[ht]
\begin{center}
\begin{tabular}{|c|c|c|c|}
\hline
\cline{2-4}
$\eta - window$&$k$&$<n>$&{$\chi^2/ndf$}\\
\hline
$2.0 < \eta < 2.5$&$1.92\pm0.02$&$3.49\pm0.03$&$195.80/18$\\
$2.5 < \eta < 3.0$&$1.98\pm0.02$&$3.40\pm0.03$&$231.59/18$\\
$3.0 < \eta < 3.5$&$2.12\pm0.02$&$3.26\pm0.03$&$228.71/18$\\
$3.5 < \eta < 4.0$&$2.35\pm0.03$&$3.08\pm0.03$&$233.34/18$\\
$4.0 < \eta < 4.5$&$2.81\pm0.05$&$2.88\pm0.03$&$240.78/18$\\
\hline
\end{tabular}
\caption{Values of parameters of NBD functions as obtained by fitting the multiplicity distributions for the primary charged particles
in minimum bias events in $pp$ - collisions at $\sqrt {s}$ = 7 TeV for five small $\eta$-windows.}
\label{tab:NBD_minbias}
\end{center}
\end{table}
As can be seen from the $\chi^2 / ndf$ values, the single NBD function is far from a satisfactory description of the multiplicity distribution for the 
minimum bias events. The fitted values of the NBD parameters, however, show consistent dependence on the position of the $\eta$ - bin. 
The average multiplicity ($<n>$) decreases and the $k$-parameter increases indicating broader distributions in the psedurapidity bins in 
the more forward regions.

\begin{center}
  \begin{figure}[h]
    \includegraphics[scale=0.38]{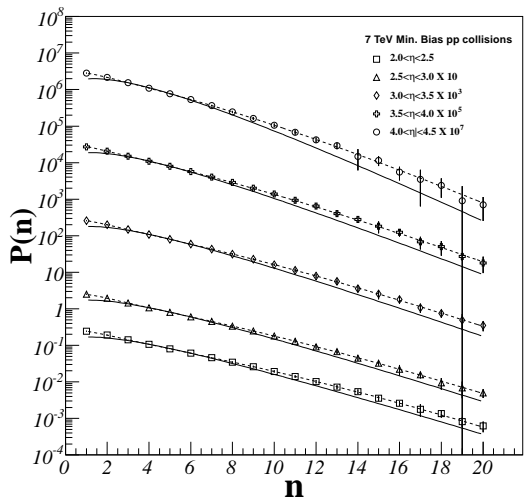} 
    \caption{Primary charged particle multiplicity distributions for minimum bias $pp$ - collisions at $\sqrt {s}$ = 7 TeV for different  $\eta$-windows 
    of width $\eta_{c} = 0.5$ scanning over the $\eta$ - range $2.0 < \eta < 4.5$. The solid lines drawn along the data-points correspond to respective 
    fits of a single NBD, while the dotted lines represent the two-NBD fits. The error-bars include both the statistical and the systematic uncertainties.}
    \label{fig:minbias_smallbins} 
  \end{figure}
\end{center}

The multiplicity distributions for the minimum bias events could be better described by the weighted superposition of two NBDs as can be seen from the 
plots in Fig.~\ref{fig:minbias_smallbins}, where the multiplicity distributions along with the single NBD and the two-NBD fits have been plotted. The 
deviation for the single NBD fits is more for the distributions in the more forward region. Though from the plots in the Fig.~\ref{fig:minbias_smallbins} 
and the $\chi^2 / ndf$, as listed in the Table - ~\ref{tab:twoNBD_minbias}, it is clear that two-NBD describes the minimum bias data better, the values 
of the fit parameters with large errors in these small rapidity intervals, as tabulated, are not suitable to reveal systematic behavior of the parameters. 
At this point, we recollect that the multiplicity distributions of the charged hadrons produced in Non-Single Diffractive (NSD) events of $pp$ collisions 
at $\sqrt {s}$ = 7 TeV \cite {ref04} in overlapping pseudorapidity bins of different widths, $|\eta|$ = $\eta_{c}$ = 0.5 to 2.4, also fit better \cite {ref13}  to 
the two-NBD than a single NBD function. It is worth mentioning, however, that the Clan-structure description of Ref.- \cite{ref09} failed to match the 
mid-rapidity LHC data \cite {ref12, ref13}.  
\begin{table}[h]
\begin{center}
\begin{tabular}{|c|c|c|c|c|}
\hline
$k_{soft}$&$\langle n \rangle_{soft}$&$k_{semihard}$&$\langle n \rangle_{semihard}$&{$\chi^2/ndf$}\\
\hline
$3.15\pm2.56$&$5.14\pm2.06$&$1.92\pm0.96$&$1.53\pm0.80$&$0.70/14$\\
$2.72\pm2.23$&$4.54\pm1.43$&$2.31\pm2.25$&$1.32\pm0.54$&$0.47/14$\\
$3.18\pm1.65$&$4.63\pm2.01$&$2.05\pm1.19$&$1.43\pm0.83$&$0.31/14$\\
$1.98\pm0.54$&$1.53\pm0.21$&$3.99\pm2.99$&$4.73\pm0.35$&$0.29/14$\\
$2.11\pm0.95$&$1.30\pm0.47$&$3.55\pm0.89$&$3.93\pm0.80$&$0.39/14$\\
\hline
\end{tabular}
\caption{Values of parameters of Two-NBD as obtained by fitting the multiplicity distributions for the primary charged particles
in minimum bias events in $pp$ - collisions at $\sqrt {s}$ = 7 TeV for five $\eta$-windows, tabulated in the same order as in 
Table - ~\ref{tab:NBD_minbias}.}
\label{tab:twoNBD_minbias}
\end{center}
\end{table}

In the context of the hard-QCD events, it may be noted that the LHCb experiment selected an event with at least one particle with transverse 
momentum greater than 1 GeV/c in the range 2.5 $< \eta <$4.5, as a "hard" event. Similar approach was adopted \cite {ref16} by the CDF experiment 
at Tevatron, Fermilab where two isolated sub-samples, soft and hard, were analyzed separately to reveal that the properties of the soft sample were 
invariant as a function of the centre-of-mass energy. The CDF experiment isolated events considering the events with no particle of transverse energy, 
$E_{T} > $ 1.1 GeV as "soft" event. Though, none of the other experiments of LHC has measured multiplicity distribution for the so-called hard-QCD events, 
the invariance of multiplicity distribution of soft events as a function of $\sqrt {s}$ has been revealed \cite {ref13} in the analysis of the data of the CMS 
experiment \cite {ref04} in terms of two-NBD. Considering that the two-component model of particle productions is valid in the forward region and that the 
criterion for isolating the hard-QCD events is proper, one may expect the multiplicity distributions for the hard-QCD events to be well described 
by a single NBD function only. 
\begin{center}
  \begin{figure}[h]
    \includegraphics[scale=0.38]{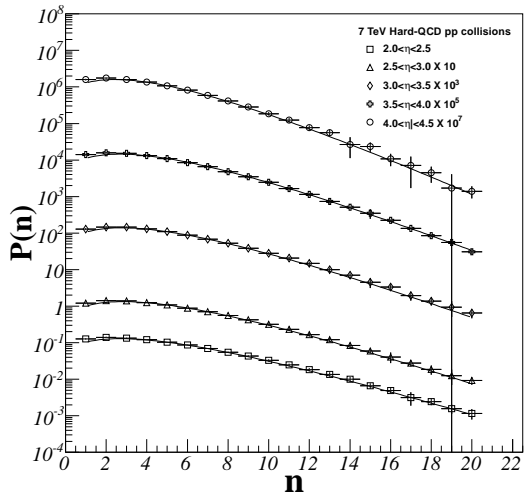} 
        \caption{Primary charged particle multiplicity distributions for the hard-QCD events in $pp$ - collisions at $\sqrt {s}$ = 7 TeV for different  $\eta$-windows 
    of width $\eta_{c} = 0.5$ scanning over the $\eta$ - range $2.0 < \eta < 4.5$. The solid lines drawn along the data-points correspond to respective 
    fits of a single NBD. The error-bars include both the statistical and the systematic uncertainties.}
    \label{fig:hard_smallbins} 
  \end{figure}
\end{center}
We fit a single NBD function to the multiplicity distributions of the produced primary charged particles for the hard-QCD events of LHCb experiments 
\cite {ref05} in small non-overlapping pseudo rapidity bins. The relevant plots are depicted in the Fig.~\ref{fig:hard_smallbins}. The plots in the 
Fig.~\ref{fig:hard_smallbins} show that the single NBD function fits reasonably well to the multiplicity distributions in small $\eta$ - windows. 
The values of $\chi^2/ndf$ for the respective plots are given in Table - ~\ref{tab:NBD_hard}. For two of $\eta$ - windows, however, the values of  
$\chi^2/ndf$ are not satisfactory. The values of the parameters $<n>$ and $k$, as tabulated in Table - ~\ref{tab:NBD_hard} show systematic trend, the
$<n>$ decreases and the $k$ increases with shift of the $\eta$-window more towards forward rapidity. On the whole, the single NBD appears to
describe the multiplicity distributions for the hard-QCD events in small $\eta$-windows in the forward region.
\begin{table}[ht]
\begin{center}
\begin{tabular}{|c|c|c|c|}
\hline
$\eta - window$&$k$&$<n>$&{$\chi^2/ndf$}\\
\hline
$2.0 < \eta < 2.5$& $2.83\pm0.05$& $4.95\pm0.05$&$29.14/18$\\
$2.5 < \eta < 3.0$&$3.19\pm0.04$&$4.86\pm0.05$&$23.59/18$\\
$3.0 < \eta < 3.5$&$3.39\pm0.05$&$4.64\pm0.05$&$87.69/18$\\
$3.5 < \eta < 4.0$&$3.53\pm0.33$&$4.45\pm0.02$&$45.34/18$\\
$4.0 < \eta < 4.5$&$3.82\pm0.06$&$3.97\pm0.04$&$27.47/18$\\
\hline
\end{tabular}
\caption{Values of parameters of NBD functions as obtained by fitting the multiplicity distributions for the primary charged particles
in the hard-QCD events in $pp$ - collisions at $\sqrt {s}$ = 7 TeV for five $\eta$-windows of width $\eta_{c} = 0.5$, each.} 
\label{tab:NBD_hard}
\end{center}
\end{table}

\begin{center}
\begin{figure}[h]
\includegraphics[scale=0.38]{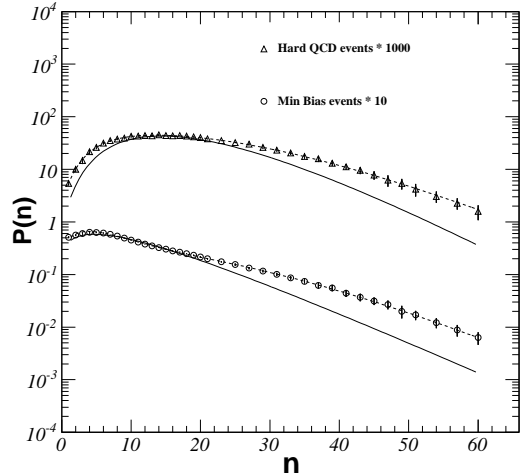} 
    \caption{Primary charged particle multiplicity distributions for the minimum-bias and the hard-QCD events in $pp$ - collisions at $\sqrt {s}$ = 7 TeV for 
    the $\eta$-window of width $\eta_{c} = 2.5$ ($2.0 < \eta < 4.5$). The solid lines drawn along the data-points correspond to respective 
    fits of NBD, while the dotted lines represent the two-NBD fits. The error-bars include both the statistical and the systematic uncertainties.}
    \label{fig:compare_largebins} 
\end{figure}
\end{center}
We continue to fit the NBD function to the multiplicity distributions for the wider pseudo rapidity range, $\eta_{c} = 2.5$ in the $\eta$ - range 
$2.0 < \eta < 4.5$ for both the event classes. As can be seen from the $\chi^2/ndf$ values obtained from the best-fit methods and tabulated in the
Tables - ~\ref{tab:NBD_largebin}, both the distributions do not fit a single NBD function. This led us to the consideration of the weighted superposition 
of two-NBD in describing both the event-classes. In case of hard-QCD events, of course, the terminology of the two-component model in respect of the 
Eq. - 3 becomes irrelevant and it is just the functional form of the equation that we are interested in. In the Tables - ~\ref{tab:twoNBD_largebin}, we 
denote the two components of the multiplicity distribution of the hard-QCD events with suffixes 1 and 2. 
\begin{table}[h]
\begin{center}
\label{tab}
\begin{tabular}{|c|c|c|c|}
\hline
$Event-class$&$k$&$<n>$&{$\chi^2/ndf$}\\
\hline
$Min. Bias$&$1.81\pm0.01$&$11.63\pm0.06$&$853.78/37$\\
\hline
$Hard-QCD$&$4.32\pm0.08$& $19.35\pm0.15$&$559.65/37$ \\
\hline
\end{tabular}
\caption{Values of parameters of the single NBD function as obtained by fitting the multiplicity distributions for the 
primary charged particles in $pp$ - collisions at $\sqrt {s}$ = 7 TeV for $\eta$-window of width $\eta_{c} = 2.5$ ($2.0 < \eta < 4.5$).}
\label{tab:NBD_largebin}
\end{center}
\end{table}
The Fig.~\ref{fig:compare_largebins} depicts the primary charged particle multiplicity distributions for the minimum-bias and the hard-QCD events 
along with corresponding best fits with a single NBD function and with the superposition of two-NBDs. The values of the fit parameters, obtained by the 
best fits in terms of $\chi^2/ndf$, as well as the $\chi^2/ndf$ values are tabulated in the Tables - ~\ref{tab:twoNBD_largebin}, for the two-NBD. 
The Fig.~\ref{fig:compare_largebins} and the $\chi^2/ndf$ values tabulated in Tables - ~\ref{tab:NBD_largebin} and ~\ref{tab:twoNBD_largebin} clearly 
indicate that the multiplicity distributions for both the minimum bias and the the so-called hard-QCD events, indeed are better described by the two-NBD 
than a single NBD function  for the large pseudorapidity bin, $\eta_{c} = 2.5$ in the $\eta$ - range $2.0 < \eta < 4.5$.  

\begin{table}[h]
\begin{center}
\label{tab}
\begin{tabular}{|c|c|c|c|c|}
\hline
$k_{soft(1)}$&$\langle n \rangle_{soft(1)}$&$k_{semihard(2)}$&$\langle n \rangle_{semihard(2)}$&{$\chi^2/ndf$}\\
\hline
$2.23\pm0.15$&$7.30\pm0.75$&$4.11\pm1.00$&$23.38\pm2.04$&$16.31/33$\\
\hline
$4.04\pm0.62$&$10.64\pm1.86$&$4.20\pm0.85$&$24.47\pm1.42$&$4.62/33$ \\
\hline
\end{tabular}
\caption{Values of parameters of the Two-NBD as obtained by fitting the multiplicity distributions for the 
primary charged particles in $pp$ - collisions at $\sqrt {s}$ = 7 TeV for $\eta$-window of width $\eta_{c} = 2.5$ ($2.0 < \eta < 4.5$), for two 
classes of events, the minimum-bias and the hard-QCD, tabulated in the same order as in Table - ~\ref{tab:NBD_largebin}.}
\label{tab:twoNBD_largebin}
\end{center}
\end{table}
\section{Summary and Remarks}
\label{}

We have analyzed the multiplicity distributions of primary charged particles in $pp$ collisions at $\sqrt {s}$ = 7 TeV, as measured by the 
LHCb experiment at the LHC. The LHCb has measured the multiplicity distributions in several small ($\Delta\eta <$ 0.5) pseudorapidity windows 
mostly in the forward $\eta$ - region, $2.0 < \eta < 4.5$, as well as in a large $\eta$ - window ($\Delta\eta <$ 2.5) for two
classes of events, the minimum bias and the hard-QCD. The distributions have been analyzed in terms of the NBD function.

For the minimum bias events, we observe that the distributions in both the small and the large pseudorapidity windows could be better described 
by the weighted superposition of two-NBDs than a single NBD function - a feature similar to what has been exhibited by the multiplicity distributions 
of primary charged hadrons produced in the mid-rapidity region in the $pp$ collisions at $\sqrt {s}$ = 7 TeV. 

The reasonable good fits of the single NBD to the multiplicity distributions for the "hard" events in small $\eta$ - windows also are more-or-less in 
agreement with the two component model of the so-called "hard" and "soft" particle productions. But, the need of a similar function formed by the 
weighted superposition of two NBDs in describing multiplicity distribution of the "hard" events in the large $\eta$ - window contradicts the concept 
of the discussed two-component model. 

On the basis of the finding that the multiplicity distribution of "hard" events in large $\eta$ - window deviates appreciably from a single NBD and 
requires weighted superposition of two NBDs, one may conclude that the discussed two-component model \cite {ref09} does not conform 
fully with the multiplicity distribution in the forward-rapidity region of $pp$ collisions at $\sqrt {s}$ = 7 TeV. The finding could be attributed either to 
biased selection criterion of the 'hard events' or to the possibility of different particle production mechanism in different phase space. 

It is worth discussing at this point that there exists no specific orthogonal variable, as yet, to separate the "soft" and the "hard" events 
in $pp$ collisions. Isolating the "hard" ("soft") events on the basis of having at least one (no) particle with the transverse momentum or transverse 
energy greater than a certain given value is a data driven approach and may inherit some biases, which need corrections. The selection criterion 
of the hard interaction events at LHCb resulted the geometrical acceptance no longer independent of momentum and the distributions were 
accordingly corrected by the collaboration \cite {ref06}. In this study, we have analyzed the corrected distributions.

To conclude on the possibility of different particle production mechanism at different phase space, direct comparison of similar analysis in the 
mid-rapidity and in the forward-rapidity is essential. Our study with the minimum bias events at the forward rapidity and comparison of the related 
results with similar study with the mid-rapidity data \cite {ref13} do not indicate to the possibility of different particle production mechanism,
in the framework of the two-component model. The results of our analysis with the 'hard-QCD' events, on the other hand, could not be compared with
the mid-rapidity data as there exists no measured multiplicity distribution for "hard" events in the mid-rapidity region at LHC. In the present scenario, 
similar analysis of isolated "hard" events of $pp$ collisions in the mid-rapidity region would be useful to obtain a comprehensive picture on the role 
of the discussed two-component model vis-a-vis the multiplicity distributions in different phase space in $pp$ collisions at the LHC energies. 

Also, other theoretical and phenomenological approaches \cite {ref15, ref17, ref18}, successful in describing the data at the mid-rapidity, may 
be compared with the LHCb data at the forward rapidity region.  In reference \cite {ref17}, the CMS data \cite {ref04} at the mid-rapidity have been 
successfully described in the framework of Independent Pair Parton Interaction (IPPI) \cite{ref19} and in terms of Quark Gluon String Model (QGSM) 
\cite{ref20,ref21} that fits better to the data than the IPPI model. The mid-rapidity data have been analyzed \cite {ref22} also in the light of another 
multiple scattering model of particle production, the Dual Parton Model (DPM).

Note: During the review process of this article, we came across an article \cite {ref23} that reports analysis of minimum bias multiplicity distributions 
measured by all the experiments at the LHC by weighted superposition of three NBD functions.

\end{document}